\documentclass[a4paper]{article}
\usepackage{jinstpub}

 \usepackage{siunitx}
 \usepackage{subfig}
\usepackage{lineno}
\usepackage[colorinlistoftodos]{todonotes}
\title{Acceptance tests of Hamamatsu R7081 photomultiplier tubes}
\author[a]{O. A. Akindele,}
\author[a]{A. Bernstein,}
\author[b]{S. Boyd,}
\author[c]{J.~Burns,}
\author[c,d]{M.~Calle,}
\author[e]{J.~Coleman,}
\author[e]{R.~Collins,}
\author[d]{A.~C.~Ezeribe,}
\author[f]{J. He,}
\author[e]{G.~Holt,}
\author[b]{K.~Jewkes,}
\author[d]{R.~Jones,}
\author[d,1]{L.~Kneale\note{Corresponding author.},}
\author[c]{P.~Lewis,}
\author[d]{M.~Malek,}
\author[g]{C. Mauger,}
\author[b]{A.~Mitra,}
\author[h]{F.~Muheim,}
\author[h,1]{M.~Needham,}
\author[i]{S.~M.~Paling,}
\author[f,2]{L. Pickard\note{Now at University of California, Berkeley, CA 94720.},}
\author[c]{S.~Quillin,}
\author[d]{J.~Rex,}
\author[i]{P.~R.~Scovell,}
\author[c,d]{T.~Shaw,}
\author[h]{G.~D.~Smith,}
\author[g]{G. Soos,}
\author[d,i]{C.~Toth,}
\author[b,3]{S.~Valder\note{Now at Rutherford Appleton Laboratory, Didcot OX11 0DE.},}
\author[h]{B.~Wade,}
\author[d]{H.~Willett,}
\author[d]{S.~Wilson}
\affiliation[a]{\small Lawrence Livermore National Laboratory, 7000 East Ave., Livermore, CA 94550, USA}
\affiliation[b]{\small University of Warwick, Coventry, West Midlands CV4 7AL, UK}
\affiliation[c]{\small Atomic Weapons Establishment, Aldermaston, Reading RG7 4PR, UK}
\affiliation[d]{\small University of Sheffield, Hounsfield Road S3 7RH, UK}
\affiliation[e]{\small University of Liverpool, Oxford Street, Liverpool L69 7ZE, UK}
\affiliation[f]{University of California, Davis, One Shields Avenue, Davis, CA 95616, USA}
\affiliation[g]{University of Pennsylvania, 209 South 33rd Street, Philadelphia, PA 19104, USA}
\affiliation[h]{\small University of Edinburgh, Mayfield Road, EH9 3JZ, UK}
\affiliation[i]{\small STFC Boulby Underground Laboratory, Saltburn-by-the-Sea TS13 4UZ, UK}

\emailAdd{e.kneale@sheffield.ac.uk}
\emailAdd{matthew.needham@ed.ac.uk}
\abstract{Photomultiplier tubes (PMTs) are traditionally an integral part of large underground experiments as they measure the light emission from particle interactions within the enclosed detection media. The BUTTON experiment will utilise around 100 PMTs to measure the response of different media suitable for rare event searches. A subset of low-radioactivity 10-inch Hamamatsu R7081 PMTs were tested, characterised, and compared to manufacture certification. This manuscript describes the laboratory tests and analysis of gain, peak-to-valley ratio and dark rate of the PMTs to give an understanding of the charge response, signal-to-noise ratio and dark noise background as an acceptance test of the suitability of these PMTs for water-based detectors. Following the evaluation of these tests, the PMT performance agreed with the manufacturer specifications. These results are imperative for modeling the PMT response in detector simulations and providing confidence in the performance of the devices once installed in the detector underground.}
\begin{document}
%
%
%
\maketitle
\flushbottom
%
%

\section{Introduction}
\label{sec:intro}
Precise single-photon detection is essential for the reconstruction of the products of neutrino interactions in water Cherenkov detectors. The WATCHMAN collaboration~\cite{Askins:2015bmb,Bernstein2019} proposed to build a gadolinium-loaded water Cherenkov detector to demonstrate the use of antineutrinos as a tool for nuclear non-proliferation via remote reactor monitoring. The chosen site was the STFC Boulby Underground Laboratory facility which is located around 25~km from the EDF Hartlepool nuclear reactor complex~\cite{Akindele2023}. The early closure of the Hartlepool reactors led to the cancellation of the project at Boulby in 2022. However, a 30-tonne demonstrator, BUTTON (Boulby Underground Technology Testbed for Observing Neutrinos) detector, is being designed to test technologies for an antineutrino detector for remote monitoring of nuclear reactors. The photon detection system of BUTTON will use  around one hundred 10 inch (253 mm) Hamamatsu R7081 photomultiplier tubes (PMTs) with low radioactivity glass~\cite{R7081datasheet} that 
were originally purchased as a pre-series by the WATCHMAN collaboration. In this paper, electrical tests of 87 of these PMTs are reported. The aim of these tests was to demonstrate that these PMTs met the requirements of a large-scale detector and to prepare for the series production quality assurance. Since the R7081 PMT series is widely used the results here can be compared with previous studies~\cite{Bauer:2011ne,AGUILAR2005132,Abbasi2010} as well as the shipping data provided by Hamamatsu.

The setup used in the pre-series testing is described in section \ref{sec:setup} and the data taking, processing and quality are described in section \ref{sec:processing}. The analysis and results are described in sections \ref{sec:gain}, \ref{sec:peak} and \ref{sec:dark} and the results are summarised in section \ref{sec:summary}.

\section{Setup}
\label{sec:setup}
\begin{figure}[htbp]
\begin{center}
\includegraphics[width=4.3 cm]{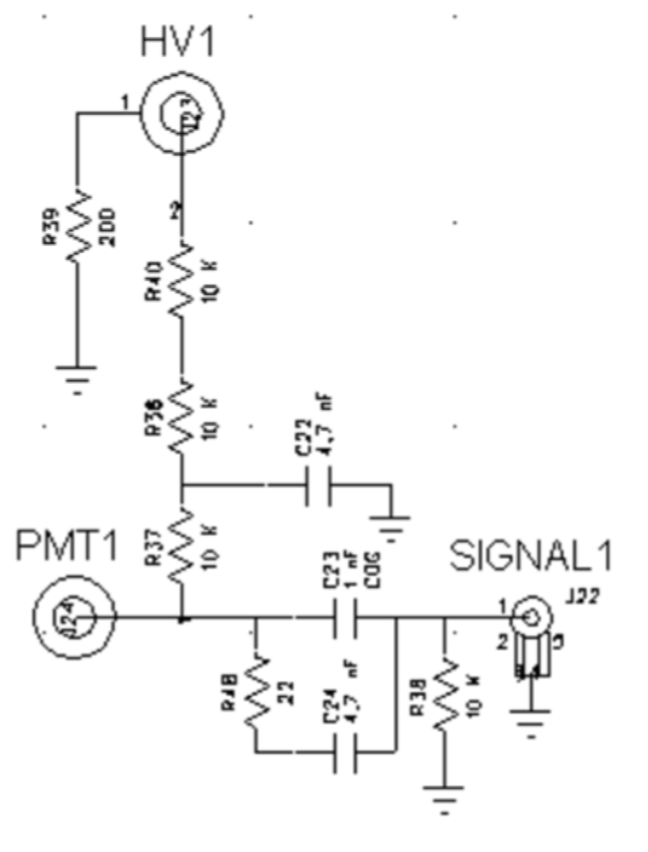}
\caption{Splitter box channel design to separate the signal output from the PMT and the HV supply to the PMT.}
\label{fig:splitter}
\end{center}
\end{figure}

%
Each PMT is supplied with an integrated base provided by Hamamatsu and has the dynode chain given in table \ref{tab:divider_ratios}. The base is encapsulated and tested to be compatible for use in Gd water. This is attached to an 80~m long 50~$\Omega$ BELDEN YR53485 cable which supplies the PMT with positive high voltage (HV) and provides the signal return path. The signal is separated from the HV in a custom-built four-channel splitter box based on the design shown in figure~\ref{fig:splitter}.
\begin{table}[htbp]
  \caption{Ratios for the voltage divider chain used for biasing the R7081 PMT~\cite{Hamamatsu2018}. \\
  K = Cathode, D = Dynode, P = Anode, F = Focus\label{tab:divider_ratios}}
\begin{tabular}{l|l|l|l|l|l|l|l|l|l|l|l|l|l|l|l}
    \hline
    Electrodes & K & D1 & F2 & F1 & F3 & D2 & D3 & D4 & D5 & D6 & D7 & D8 & D9 & D10 & P\\
    \hline
    Ratio & - & 16.8 & 0 & 0.6 & 0 & 3.4 & 5 & 3.33 & 1.67 & 1 & 1.2 & 1.5 & 2.2 & 3 & 2.4 \\
    \hline
    \end{tabular}
\end{table}

\begin{figure}[htbp]
\begin{center}
\includegraphics[width=8.6 cm]{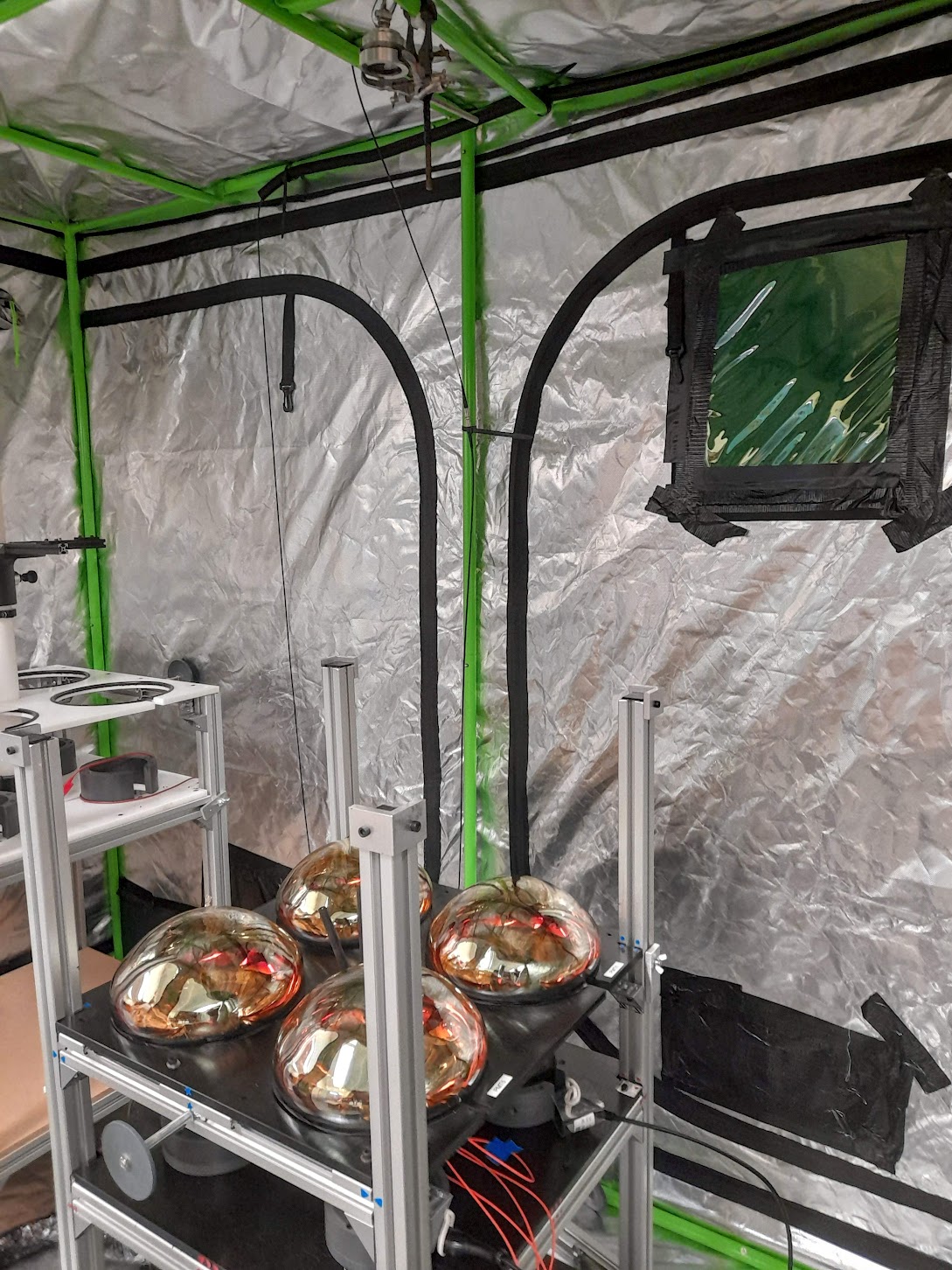}
\caption{Mounting of the PMTs in the mechanical rig in the tent. Optical fibres guide the light from the LED box outside the tent to the diffuser, which hangs from a fixation point on the tent roof at the top centre of the image.}
\label{fig:rig}
\end{center}
\end{figure}

The setup used for testing is shown in figure~\ref{fig:rig}. The tests were carried out in a commercial `grow' tent, which is lined with aluminised Mylar. The tent acts as a dark box and also provides electromagnetic shielding. Further stray light-level reduction was achieved using a tailored cloth cover placed over the entire the tent. In the tent, four PMTs were mounted on a custom-built rig. A 470~nm pulsed LED was used as a light source. This was located in a custom driver unit which was triggered at 10 kHz to generate pulses of light, each a few nanoseconds in length. Inside the LED box, the light from the LED is directed via a groove onto the back of optical fibres, which guide the light into a purpose-constructed dark enclosure housing the PMTs under test. The intensity of light was controlled using the LED driver voltage which was set so that the PMT signal spectra were dominated by single-photoelectron (SPE) events. To give uniform illumination of the PMT surface, the optical fibre was connected to a polymethyl methacrylate (PMMA) diffuser~\cite{Valder2018}. The mean number of photoelectrons determined from the observed charged spectra was in the range 0.1--0.2. After mounting, the PMTs were left for 18 hours under bias voltage to allow the dark rate to stabilize. The temperature of the room was measured to be in the range 21--23\textdegree C during the tests.

\begin{figure}[htbp]
\begin{center}
\includegraphics[width = 8.6 cm]{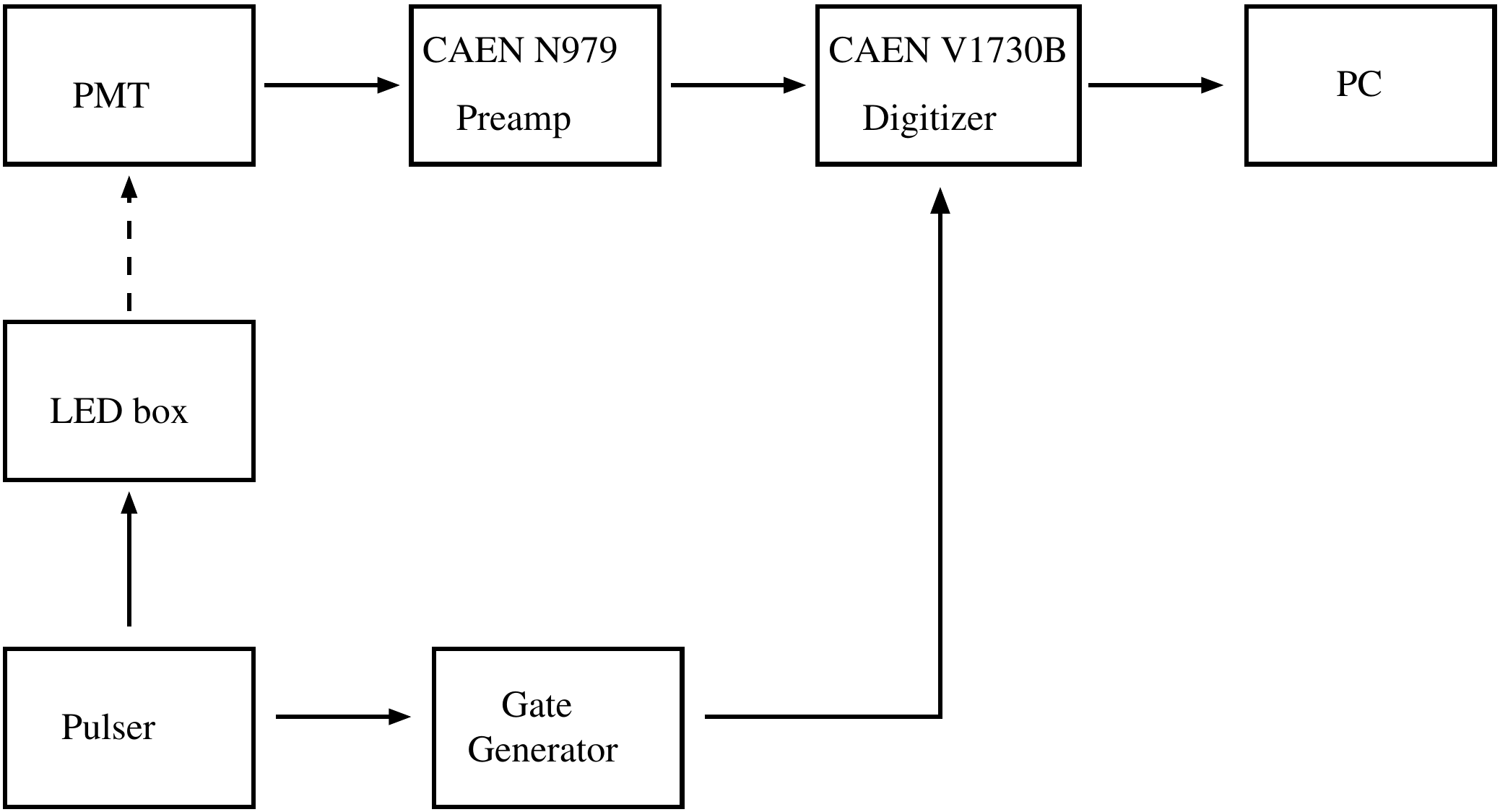}
\caption{Electronic readout scheme used for the tests. The digitiser is triggered by a delayed copy of the LED trigger from the gate generator.}
\label{fig:readout}
\end{center}
\end{figure}

The electronic readout scheme is shown in figure~\ref{fig:readout}. Commercially available electronics were used for readout. Signals from the PMTs were amplified by a factor of 10 using a CAEN N979 unit and then digitized using a CAEN V1730B module that samples at 500~MHz rate. The system was triggered by a delayed copy of the main pulser signal used to trigger the LED driver. The digitiser and LED were triggered at 10~kHz and there was no dead time. Sampled waveforms were transmitted via an optical link to a PC and stored for further analysis. Since the SPE signals being detected were small, care was taken in system grounding and isolation to reduce signal noise. For example, ferrite beads were attached to all power leads and isolators were used on USB cables required in the setup for HV control and computing accessories.

\section{Data taking, processing and quality}
\label{sec:processing}

\begin{figure}
\centering
\includegraphics[width=8.6cm]{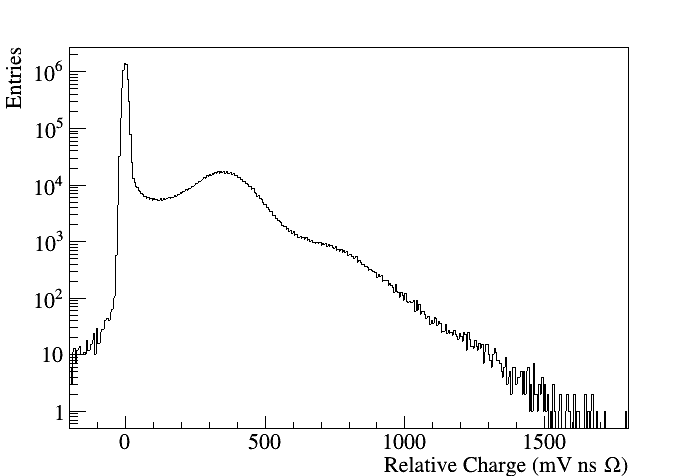}
\caption{Typical relative charge spectrum with a pedestal peak at $\sim$0 mV~ns~$\Omega$, a single-photoelectron peak at $\sim$400 mV~ns~$\Omega$ and later multiple-photoelectron peaks in the tail.}
\label{fig:ChargeSpectrum}
\end{figure}

Table \ref{tab:tests} lists details of the four PMT performance characterisation tests that were performed. Data were acquired for all of these tests for every PMT at least once and used to analyse the gain, peak-to-valley ratio and dark rate of the PMTs. 
\begin{table}[htbp]
  \begin{center}
    \caption{PMT test information. Nominal HV is the manufacturer's stated HV for $10^{7}$ gain. SPE refers to data taken using an LED light source with its intensity set to yield a single photo-electron (SPE) level at the PMT photocathode. The digitiser and LED trigger rates were set at 10 kHz for all tests. \label{tab:tests}}
    \begin{tabular}{ l |   l |     l    l       l |      l} 
    \hline
          {} & {Test Type} &   {Light pulses}  & {Duration}    & {Gate} &  {Description}   \\
          {} &             &   {(M)}     & {(mins)}  & {(ns)}     &  \\
      \hline
      1    & Nominal HV               		      & 3.0     	         & 5.0 	 & 220 		  		 & SPE at nominal HV  \\
      2    & Gain 			      & 3.0  	   	  & 5.0    & 220   		   		& SPE, repeated for 5 HV steps  \\
      3    & Peak-to-valley  				      & 3.0       		  & 5.0	  & 220    		   		& SPE at $10^{7}$ gain \\
      4    & Dark Counts      			  	      & 9.0     		 & 15.0   	  & 220    		  		& No LED \\                  
      \hline
    \end{tabular}
  \end{center}
\end{table}
Data were recorded as binary files using CAEN's \emph{WaveDump} software~\cite{WaveDump} and then analysed using a ROOT-based framework~\cite{BRUN1997}. Histograms of the charge output at the PMT anode (see for instance figure \ref{fig:ChargeSpectrum}) were created by integrating in a 50 ns time window around the SPE peak. These were used for the gain and peak-to-valley tests and are described in more detail in sections~\ref{sec:gain}~and~\ref{sec:peak}.

The signal integration window was chosen to start 15~ns before the observed mean of the coincidence time peak to ensure the rising edge and tail of the pulses were included, and to accommodate trigger jitter. The width ($\sigma$) of the timing peak fit was found to be of order 8~ns, consistent with the timing characteristics of the LED, clock jitter and the transition time spread of the PMT. This method of signal integration was developed to accommodate fast bulk data monitoring with a single routine.

%

%
\subsection{Data Quality}
\label{subsec:dataquality}
The quality of data was monitored throughout data taking. Though the system was well grounded and isolated, several sources of noise were visible. From Fourier transforms of the waveforms, narrow high frequency signals were seen that were consistent with the known frequencies used for USB protocol transport in the setup and a local radio transmitter. To handle the remaining system noise, a technique for the rejection of noisy waveforms was developed, whereby any waveform with a maximum and minimum within a factor of 2 was classed as radio-frequency noise and the waveform was rejected. Fewer than 1 - 2\% of waveforms were rejected in this way.

\section{Gain}
\label{sec:gain}
The PMT gain $\mathcal{A}$ is the ratio of the charge output at the anode to the number of photoelectrons produced by conversion at the photocathode. Gain calibration ensures the correct interpretation of the charge output at the PMT anode and is essential in setting operating voltages to achieve a uniform charge response across all PMTs in the detector.

The gain tests verify that the required $10^{7}$ gain can be achieved within the  R7801 operating voltage range of 0-2000V. A PMT would be rejected if the required  gain cannot be achieved within this range.

\subsection{Calculation of the operating voltage}

The total gain across the complete dynode chain (in the case of a rather linear voltage divider) can be modelled as
\begin{equation}
\mathcal{A} = a^n \left(\frac{V}{n} \right)^{n\alpha },
\label{eq:gain_power}
\end{equation}
where $V$ is the applied voltage, $n$ is the number of dynodes, and $a$ and $\alpha$ are constants specific to the PMT.\cite{Wright2017}

A gain of $10^7$ corresponds to a SPE relative charge output of 400~mV~ns~$\Omega$. The most accurate gain calculation is achieved by fitting the charge distributions produced through integration of the waveforms as discussed in section \ref{sec:processing}. Characterisation of both the charge response and inherent charge backgrounds of the PMT gives a value for the SPE relative charge output $Q_{SPE}$. The gain is then calculated from $Q_{SPE}$:
\begin{equation}
\mathcal{A} = \frac{ 2~Q_{SPE} \times 10^{-12}}{ e~f_{amp}~Z},
\label{eq:gain_calc}
\end{equation}
where $f_{amp} = 10$ is the amplifier gain, $Z = 50 \, \Omega$ is the impedance, \emph{e} is the charge of an electron and the factor two accounts for the halving of the charge due to a $50~\Omega$ AC termination inside the PMT in addition to the $50~\Omega$ termination in the CAEN N979 preamplifier shown in figure~\ref{fig:readout}.

Each PMT was tested at five voltage steps, chosen to be roughly flat in gain. The gain values obtained from the HV steps were fitted with 
 \begin{equation}
 \mathcal{A} = \left( \frac{V}{V_{opt}} \right)^{\beta},
  \label{eq:gain_curve_fit}
 \end{equation}
where $ V_{opt}$ is the operating voltage that gives $10^7$ gain and both $ V_{opt}$ and $\beta$ are free parameters\cite{Bauer:2011ne}. The operating voltage required to give $10^7$ gain was then calculated from the fit for each PMT and compared to Hamamatsu's nominal voltage for the same gain.

%
%
Features of a typical charge spectrum (figure \ref{fig:ChargeSpectrum}) can be separated into charge response (signal) and backgrounds. It has a pedestal peak, which consists of the noise inherent in a PMT when there is no signal pulse, a single-photoelectron signal peak and other, multiple-photoelectron signal peaks. A valley region between the pedestal and SPE peaks consists of events due to under-amplified signal and thermionic emission (a source of dark noise) from the dynodes. The fitting regime developed for the purposes of these tests is adapted from the method set out in~\cite{Bellamy:1994bv}.
%
The charge response has two contributions: the photoelectric conversion at the photocathode and the amplification at the dynodes.

The photoelectric conversion is the convolution of a Poisson process and a binary process, which represent the number of photons hitting the PMT and the photoelectric conversion respectively. The resulting distribution is
\begin{equation}
 P(n, \mu) = \frac{\mu^n e^{-\mu}}{n!},
\label{eq:photoelectric_conversion}
\end{equation}
where $P(n, \mu)$ is the probability that n photoelectrons will be observed when the mean number of photoelectrons collected at the first dynode is $\mu$. If the amplification is at least four per stage (and preferably $>10$), then the photoelectron peaks can be approximated with Gaussian functions:
\begin{equation}
 G(x) = \sum_{n=0}^\infty \frac{1}{\sigma_{SPE} \sqrt{2 \pi n}}\exp{ \left(-\frac{(x-nQ_{SPE})^2}{2 n \sigma_{SPE} ^2} \right)},
\label{eq:amplification}
\end{equation}
where $ \sigma_{SPE}$ is the width of the SPE distribution, x is the charge variable and $ Q_{SPE}$ is the SPE charge output.

When the amplification is low, which can be the case particularly at the first dynode, when the number of photoelectrons is only 1 or 2, the Gaussian approximation does not hold and the single and multiple-photoelectron Gaussians should be replaced with sums of Gaussians:
\begin{equation}
 G_n(x) = \sum_{m=0}^\infty \frac{(n\frac{Q_{SPE}}{Q_{SPE,2}})^m e^{-n\frac{Q_{SPE}}{Q_{SPE,2}}}}{m!} \cdot \frac{1}{\sigma_{SPE,2} \sqrt{2 \pi n}}\exp{ \left(-\frac{(x-mQ_{SPE,2})^2}{2 m \sigma_{SPE,2} ^2} \right)},
\label{eq:sumGaussians}
\end{equation} 
where $m$ is the number of electrons produced at the first dynode, $ Q_{SPE,2}$ is the charge output by a single electron emitted from the first dynode and $ \sigma_{SPE,2}$ is the width of the distribution in the Gaussian approximation.

The ideal PMT response is a convolution of the capture and photoelectric conversion with amplification:

\begin{equation}
 S_{ideal} (x) =  \sum_n P(n, \mu)  G(x).
\label{eq:charge_response}
\end{equation}
%
There are two distinct background distributions. Low-charge processes which occur in the absence of an incident photon give rise to the pedestal, which has a Gaussian distribution. Low-gain events with an incident photon, e.g. due to photoemission from the dynodes or photoelectrons which miss the first dynode, give rise to the valley. This can be approximated by an exponential~\cite{Bellamy:1994bv}. The backgrounds are the sum of these processes:
\begin{equation}
 B(x) = P(0,\mu)\frac{1}{\sigma_0  \sqrt{2 \pi}} \exp \left(- \frac{x^2}{2 \sigma_0^2} \right) + P_{exp} \theta(x) \alpha \exp (- \alpha x),
\label{eq:backgrounds}
\end{equation}
where $ P(0,\mu)$ is the Poisson probability that zero photoelectrons are produced, $\sigma_0$ is the width of the pedestal, $ P_{exp}$ is the probability that the second type of background is present and $\alpha$ is the coefficient of the exponential decrease of the valley. The condition
$\theta = 
\begin{cases}
      0 & x \leq 0\\
      1 & x>0
    \end{cases}    
$ 
ensures that there is an exponential component only if $x>0$.

Finally, the total charge distribution is modelled as
\begin{equation}
 Q(x) = P(0,\mu) G(x)_{ped}  + P_{exp} Exp_{val} + \sum_{n=0}^\infty P_{n} G_n(x),
\label{eq:fit}
\end{equation}
where $ G_{ped}$ and $ G_{n}$ are the Gaussian fits to the pedestal and n-photoelectron peaks, the Poisson distributions give the relative fractions of each peak and $ Exp_{val}$ is the exponential fit to the valley. The third term is the signal in equation~\eqref{eq:amplification} or equation~\eqref{eq:sumGaussians} and the first two terms are $B(x)$ in equation~\eqref{eq:backgrounds}.
%
\subsection{Gain calibration}
\label{subsec:gain_calibration}
The full fit model according to equation~\eqref{eq:fit}, with double-Gaussian fit to the pedestal, Gaussian approximation to the single- and double-photoelectron peaks, and exponential approximation to the valley allows a precise fit to the pedestal and SPE peak, which are key in the calculation of the SPE charge and thus in the operating voltage determination. In some cases, the sum-of-Gaussians fit was used to improve the fit for offline analysis.

\begin{figure}
\centering
\subfloat{
		\includegraphics[width=0.48\textwidth]{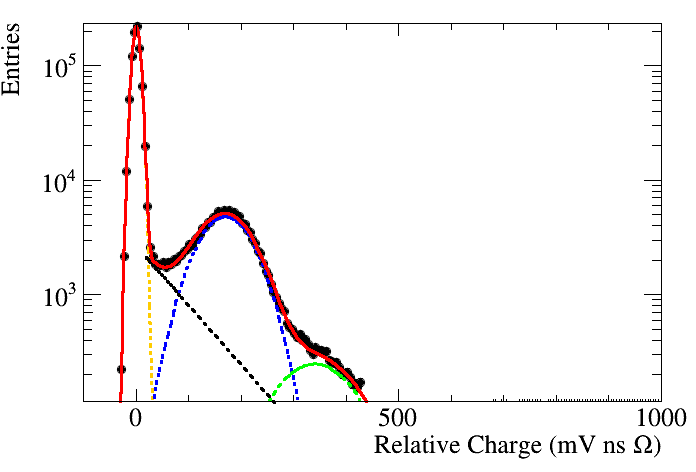}
        }
	\subfloat{	
		\includegraphics[width=0.48\textwidth]{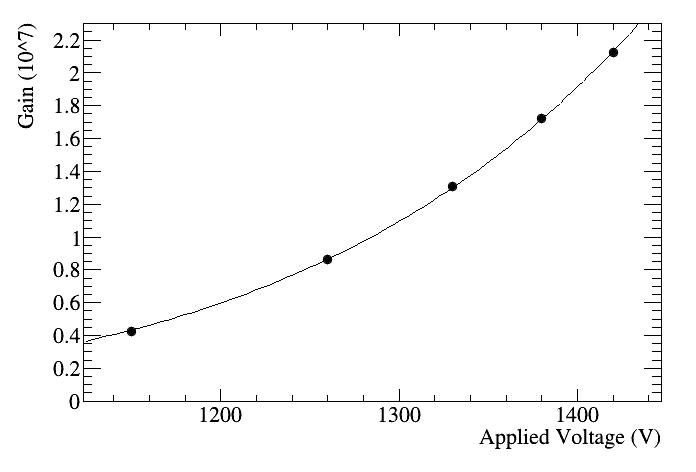}
	}
  \caption{Example of the fits for the gain calibration. The full fit to the relative charge distribution (left) shows a double-Gaussian fit to the pedestal (yellow), Gaussian approximations to the single-photoelectron (blue) and double-photoelectron (green) peaks, and an exponential approximation (black) to the valley. The fit to the gain curve (right) is achieved using five HV steps. Errors (not visible) on $Q_{spe}$ values are between 0.1 and 2.0~mV~ns~$\Omega$ and on the fit to the gain curve are < 0.0001 V}
 \label{fig:fit_final}
\end{figure}

From initial tests, a parameterisation was developed to select, for each PMT, HV values uniformly distributed in gain around the nominal operating voltage.  The length and timing of the waveforms was found to depend on the applied HV such that pulses arrived sooner and lasted for longer at higher applied voltages. The timing variation is consistent with the expectation from the data sheet~\cite{R7081datasheet} that the transit time decreases by around 12~ns from 1300~V to 2000~V. The start time and length of the integration window (section \ref{sec:processing}) was chosen to account for this effect. Figure~\ref{fig:fit_final} shows a typical fit to the SPE distribution and a typical fit of five voltage steps to extract the gain. 

The value of $\alpha$ in equation \eqref{eq:gain_power} is dependent on the gain and the voltage applied. The mean value of $\alpha$ was found to be 6.9, which agrees with a value of approximately 0.7 per stage - consistent with $10^7$ gain across 10 dynodes.

\begin{figure}[htb!]
\centering
  \includegraphics[width=8.6 cm]{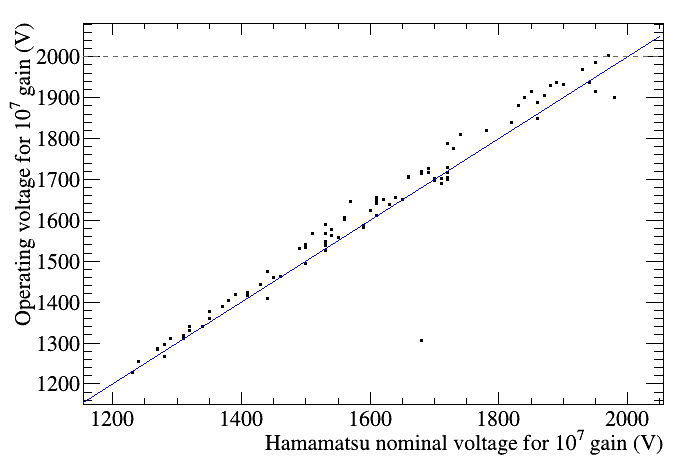}
  \caption{Comparison of calculated operating voltages with Hamamatsu nominal voltages for $10^7$ gain from full fits to filtered histograms. The grey dashed line represents the maximum operating voltage of 2000~V. The outlier has been put down to an error in the shipping data.} \label{fig:voltage_comparison_2}
\end{figure}

The results obtained are correlated with the Hamamatsu data sheet though with a factor 1.02 ($\sim 20$~V) offset (figure~\ref{fig:voltage_comparison_2}). This small difference may be due to the fact that the Hammatsu tests were made prior to the attachment of the 80~m cable which will attenuate the signal.

All but one of the PMTs were found to achieve $10^7$ gain at less than 2000~V. A single PMT had a calculated operating voltage for $10^7$ gain of 2003~V. Since all PMTs achieved the required gain at or very close to the manufacturer's recommended maximum HV, they were deemed to be acceptable. 

\section{Peak to valley}
\label{sec:peak}
The peak-to-valley, the ratio of the height of the SPE peak
to the minimum of the charge spectrum between the pedestal and SPE peak, characterises the
signal-to-noise performance of the PMT. It is determined either by fitting the full charge spectrum or by making local fits to the relevant regions. For simplicity, the second method is adopted here using the data taken at the operating voltage which gives $10^7$ gain, calculated using the method described in section \ref{sec:gain}.

The method works as follows. A local peak finding algorithm, the \emph{ROOT} class \emph{TSpectrum}, is used to identify the location of the pedestal and signal peaks. Based upon the results of this a local fit of a parabola (Gaussian) is made to the valley (signal) regions respectively. Once the locations of the minimum of the parabola and the maximum of the Gaussian are identified, the peak-to-valley ratio is straightforward to calculate. By propagating the uncertainties on the fitted parameters and by taking repeated runs, the uncertainty of the procedure is determined to be 0.01 to 0.02 (absolute uncertainty).
 
\begin{figure}[htb!]
\begin{center}
\includegraphics[width=8.6 cm]{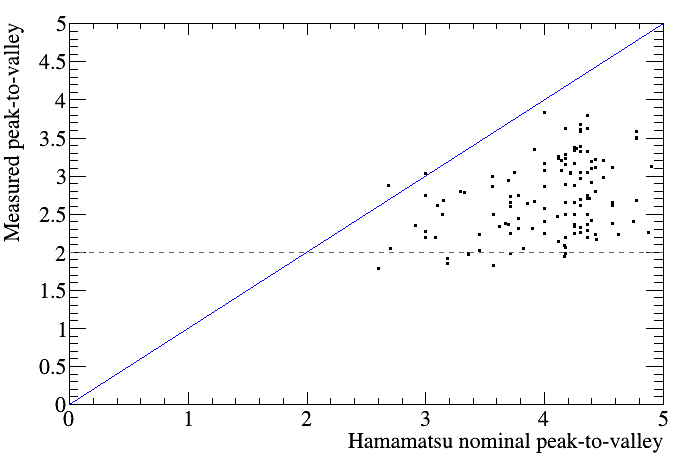}
\caption{Measured peak-to-valley measurements, versus the Hamamatsu data sheet values. The grey dashed line shows a potential requirement of a peak-to-valley ratio of 2. Errors on the values (not visible) are between 0.01-0.02 (absolute uncertainty).}
\label{fig:pv}
\end{center}
\end{figure}

Figure~\ref{fig:pv} shows the peak-to-valley ratio determined by this method, 
compared to the value provided on the Hamamatsu data sheet. Though there is a correlation between the measured peak-to-valley and the expectation from the Hamamatsu data sheet, the measured values are smaller. For a few PMTs, the peak-to-valley falls below 2, which might be considered as a reasonable criterion for rejection. Insight into the worse peak-to-valley was gained by dividing the data according to the location of the PMT in the rig. From this study, and also testing PMTs at different rig positions, it was determined that the orientation with respect to the measured magnetic field in the lab could change the peak-to-valley ratio by up to $\sim$0.5. Using magnetic shielding, as will be the case in BUTTON, all PMTs should have a peak-to-valley value larger two. An additional factor that was found to influence the measurement of the signal-to-valley is the PMT illumination, and more specifically the spatial light distribution over the photocathode, reflecting the importance of the electrical field that guides the photoelectron to the dynode chain. The diffuser resulted in values for the mean number of photoelectrons across all four PMTs which were consistent with flat illumination at the 10\% level, however this may be different to the conditions of the Hamamatsu tests.
%
%
%

\section{Dark count}
\label{sec:dark}

The dark count rate can affect event reconstruction~\cite{Li2022} so it is important to have a good understanding of the dark rates of PMTs in the detector. The requirement was that the dark count rate should be less than 10~kHz. The R7081 PMT data sheet~\cite{R7081datasheet} quotes a typical dark count after 24 hours storage in darkness of 8~kHz with a lower-level discriminator set to a threshold equivalent to one quarter of the mean SPE signal amplitude.

Dark noise pulses were identified and counted during post-processing analysis of the saved data. The tubes were biased at the calculated operating voltage in the dark tent for at least 18 hours prior to acquiring the dark rate data.

A dark pulse is deemed to have occurred in the 220~ns waveform if a signal threshold of 10~mV is exceeded. This is equivalent to $\sim$0.25 of the mean SPE signal amplitude, calculated from the peak-to-valley test data. The probability of multiple true dark pulses in a single waveform was assumed to be negligible. Multiple pulses were treated as double-pulsing phenomena and a maximum of one dark pulse per waveform was counted. 

The Hamamatsu data sheet characteristics are quoted at 25$^{\circ}$ C. For the purposes of comparison, a correction was made to the Hamamatsu nominal dark count rates to account for the difference in temperature at which the data were taken for these tests using the scaling described in section~\ref{dark:temperature}.

\begin{figure}[ht]
	\centering
	\subfloat{
		\includegraphics[width=0.48\textwidth]{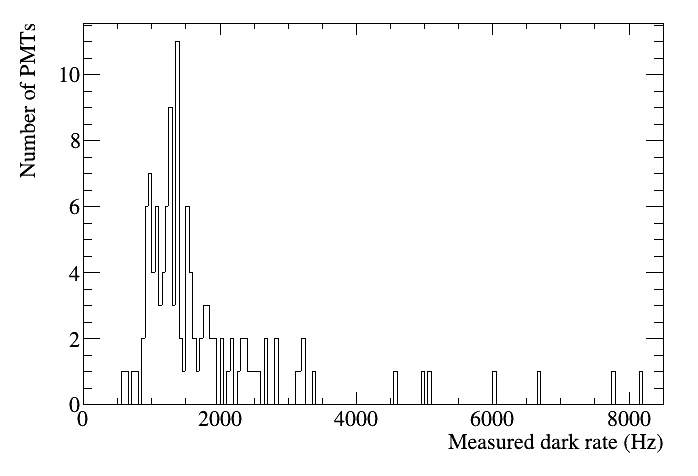}}
	\hfill
	\subfloat{	
		\includegraphics[width=0.48\textwidth]{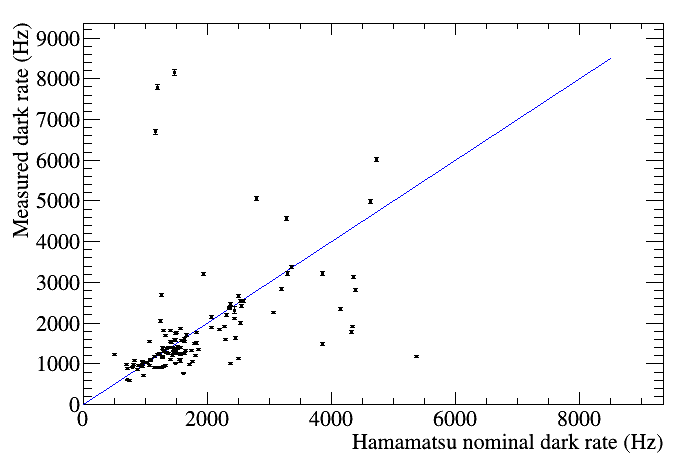}
	}
	\caption{Results of the dark rate tests showing the distribution of measured dark rate distribution in 50~Hz bins for all PMTs tested (left) and the measured dark rates compared to the Hamamatsu nominal dark rates (right).}
	\label{fig:results}
\end{figure}
All tubes showed dark rates below the 10 kHz acceptance criterion with a mean value around 1.8~kHz. For most PMTs, a good correlation with the Hamamatsu data is apparent though there are outliers in both directions.

\subsection{Dark Rate Temperature Sensitivity}
\label{dark:temperature}

\begin{figure}[ht]
	\centering
	\includegraphics[width=8.6 cm]{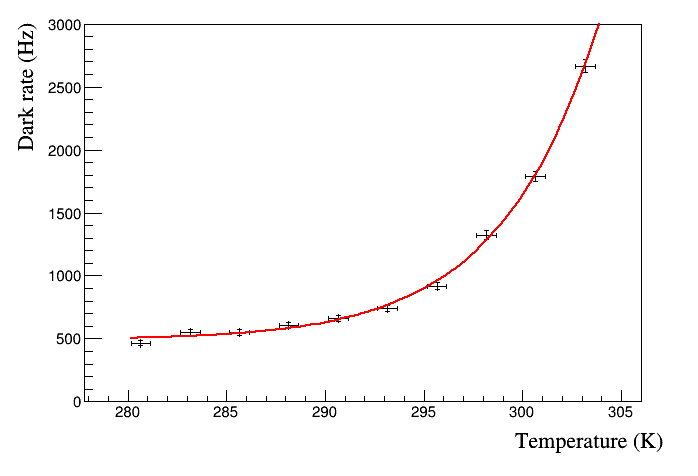}
	\caption{Dark rate versus temperature for an R7081 tube. A fit to the form described in the text is superimposed.}
	\label{fig:ratevtemp}
\end{figure}

If the dark count rate is dominated by thermionic emission, its rate should be strongly dependent on the absolute temperature \cite{FlycktMarmonier2002}. The behaviour of several tubes were measured in a climate chamber where the temperature could be controlled. Figure~\ref{fig:ratevtemp} shows the measured dark rate versus temperature for one tube. The plot suggests a temperature sensitivity of $\sim$60~Hz~K$^{-1}$ or 7\% at 20$^{\circ}$C and $\sim$150~Hz~K$^{-1}$ or 12\% at 25$^{\circ}$C. These values are consistent with the value of 100~Hz~K$^{-1}$ at 20 $^{\circ}$C reported by the Double Chooz experiment \cite{Bauer:2011ne}. 

The dark rate due to thermionic emission is expected to vary with temperature, $T$, as 
$T^{5/4} exp{(- e\psi/KT)}$, where $e$ is the electric charge, $K$ is the Boltzmann constant and $\psi$ is the work function of the photocathode \cite{hammahandbook}. The data is found to be well described by this form, with a temperature-independent offset that accounts for stray light and other non-thermal effects \cite{Wright2017}. The fitted value of the work function of the measured tubes was in the range 1.2-1.5~eV.

\section{Summary}
\label{sec:summary}
In this paper, testing of 87 PMTs for a gadolinium-loaded water-based Cherenkov detector has been described. The PMT characteristics have been shown to agree with the Hamamatsu data sheet and previous tests. All PMTs were functional and worked after transportation from Japan via the United States to the UK where they were stored in the underground Boulby laboratory. The operating voltage to achieve $10^7$ gain agrees with the Hamamatsu measurements made prior the cable attachment at the level of 20~V. Measurements of the peak-to-valley ratio were found to be generally lower than values given in the Hamamatsu data sheet and  number of PMTs were found to have a peak-to-valley below~2. The dark count measurements are reasonably well correlated with, and in general lower than, the Hamamatsu nominal dark rates, except for three PMTs with much higher dark rates than the values provided, although these were still within specifications.

\begin{acknowledgments}
This work was supported in the UK by the Atomic Weapons Establishment (AWE), as contracted by the Ministry of Defence, and the Science and Technology Facilities Council (STFC).

This work was performed under the auspices of the U.S. Department of Energy (DOE) by Lawrence Livermore National Laboratory under contract DE-AC52-07NA27344, and supported by the DOE National Nuclear Security Administration under Award Number DE-NA0000979. LLNL-JRNL-850296.

The authors would like to thank A. Papatyi for his tireless work arranging the logistics of the tests and more, and V. Li for his extensive review of the work.
\end{acknowledgments}

\newpage

\end{document}